\documentclass[aps,prb,showpacs,twocolumn,amsmath,amssymb,superscriptaddress,floatfix]{revtex4}
\usepackage{epsfig}
\usepackage{epstopdf}
\usepackage{graphicx}
\usepackage{grffile}
\usepackage{bm}
\usepackage{sidecap}
\usepackage{mathptmx}
\usepackage{mathrsfs}
\usepackage{txfonts}
\usepackage[usenames,dvipsnames,svgnames]{xcolor}
\usepackage{array}
\usepackage{float}

\usepackage{lipsum}
\DeclareMathAlphabet{\mathpzc}{OT1}{pzc}{m}{it}

\newcommand{\tobedeleted}[1]{\textcolor{green}{#1}}
\renewcommand{\tobedeleted}[1]{\relax}

\begin{document}
\renewcommand{\thefigure}{\arabic{figure}}
\def\be{\begin{equation}}
\def\ee{\end{equation}}
\def\ber{\begin{eqnarray}}
\def\eer{\end{eqnarray}}

\def\kv{{\bf k}}
\def\bfr{{\bf r}}
\def\qv{{\bf q}}
\def\pv{{\bf p}}
\def\sigmav{{\bf \sigma}}
\def\tauv{{\bf \tau}}
\newcommand{\h}[1]{{\hat {#1}}}
\newcommand{\hdg}[1]{{\hat {#1}^\dagger}}
\newcommand{\bra}[1]{\left\langle{#1}\right|}
\newcommand{\ket}[1]{\left|{#1}\right\rangle}

\title{Effect of chiral selective tunneling on quantum transport \\in magnetic topological-insulator thin films}
\date{\today}

\author {Taahere Sabze}
\affiliation{School of Physics, Damghan University, P.O. Box 36716-41167, Damghan, Iran}

\author{Hosein Cheraghchi}\email{cheraghchi@du.ac.ir}
\affiliation{School of Physics, Damghan University, P.O. Box 36716-41167, Damghan, Iran}

\begin{abstract}
The electronic transport properties in magnetically doped ultra-thin films of topological-insulators is investigated by using Landauer-buttiker formalism. The chiral selective tunneling is addressed in such systems which leads to transport gap and as a consequence current blocking. This quantum blocking of transport occurs when the magnetic states with opposite chirality are aligned energetically. This can be obsereved when an electron tunnels through a barrier or well of magnetic potential induced by the exchange field. It is proved and demonstrated that this chiral transition rule fails when structural inversion asymmetric potential or an in-plane magnetization is turning on. This new finding is useful to interpret quantum transport through topological-insulator thin films especially to shed light on longitudinal conductance behavior of quantum anomalous Hall effect. Besides, one can design electronic devices by means of magnetic topological-insulator thin films based on the chiral selective tunneling leading to negative differential resistance.
\end{abstract}
\pacs{...} \maketitle

\section{Introduction}
Topological-insulators (TI) which are new quantum state of materials has been recently paid great theoretical and experimental attentions\cite{prl95Q,prl96,prl95Z,prl97,science314,science318,nature398} for its new concepts in condensed matter physics leading to interesting phenomenon such as quantum anomalous Hall (QAH) insulators \cite{science329,science340} and also its potential applications in future electronic devices originating from high conductivity and spin-polarization of TI surface states\cite{Hassan,nature438,nature452}. As a result of strong Rashba spin-orbit interaction in these materials which gives rise a spin-momentum locking, dissipation-less edge states are gapless and protected against backscattering by time-reversal symmetry (TRS), while their bulk spectrum has a small gap\cite{Hassan,prl983D,prb76fu}. Such a Dirac-cone spectrum centered in the $\Gamma$ point was initially predicted and experimentally observed in the surface states of the Bismuth-based materials such as Bi$_2$Se$_3$ and Bi$_2$Te$_3$ namely as 3D-TIs\cite{nature452,science325}. An ultra-thin film of TI leads to reduced scattering arising from the bulk states as well as a gap opening originated from the tunneling between the top and bottom surface states when TI's thickness is thinner than 5 $nm$\cite{nature584,prb81041307}. Furthermore, in TI thin films, a topological quantum phase transition emerges from quantum spin Hall (QSH) insulator to normal insulator when structural inversion asymmetry (SIA), a potential difference between the top and bottom surfaces, exceeds a critical voltage\cite{electrically}.

The QAH effect which is spin-polarized quantized transport in the absence of external magnetic field, at first, was analytically proposed in Ref.\cite{science329} and experimentally observed in transition metals-doped of (Bi,Sb)$_2$Te$_3$ as magnetically doped topological-insulator thin films\cite{science340,prl113137201,nature10731}. Indeed, ferromagnetic ordering in these materials breaks time-reversal symmetry and induces an exchange field $M$ on TI's surfaces which essentially can change the band gap\cite{prb81H}. In enough strong exchange field ($\lvert M \rvert>\lvert \Delta_0 \rvert$), a band inversion would occur in the band structure leading to topologically nontrivial phase\cite{prb81H,electrically,prl2013QT}. Around the neutral point, the Hall conductance shows a quantized plateau originating from topologically non-trivial conduction band, where the longitudinal conductance is almost zero\cite{science340,prl113137201,nature10731}. The sudden drop of the Hall conductance is accompanied by a peak in the longitudinal conductance which as Lu. {\it et.al.}\cite{prl2013QT} showed, is attributed to the concentrated Berry curvature and also local maximum of group velocity close to the band edge, respectively. However, there are also other interpretations for the experimental results of transport through QAH insulators. To explain dissipative longitudinal conductnance, the coexistence of chiral and quasi-helical edge states in magnetic TI \cite{wang} and also in a microscopic view,  the randomly magnetic domains with opposite magnetizationsin\cite{universality} are considered. The role of magnetic disorder in suppression of transmission by the edge states accompanied with enhanced longitudinal transmission has been numerically also verified\cite{epl2014}.

 Beside transport properties, the mechanism behind strong ferromagnetic ground state in QAH insulators is, however still under debate\cite{Grauer}. One of the proposals is the existence of magnetic impurities on the surface of TI thin film which can enhance local density of state in the band gap \cite{Shiranzaee1} and as a consequence, the RKKY interaction in these materials\cite{Shiranzaee2} is intensified. On the other hand, the application of topological insulator and its thin version as the baseground for nano-electronic devices has also attracted attentions\cite{transportTI_nanoscale,transport_1,transport_2,transport_3}. Quantum transport properties through an array of electrostatic potential barriers based on non-magnetic TI thin films was studied by Li, {\it et. al.}\cite{transportTI_nanoscale} showing the effect of the gate voltage manipulation on conductance. However, in this calculation, there is no Rashba splitting in the band spectrum which is induced by SIA.      
\begin{figure}[!h]
\centering
\includegraphics[width=1.0\linewidth]{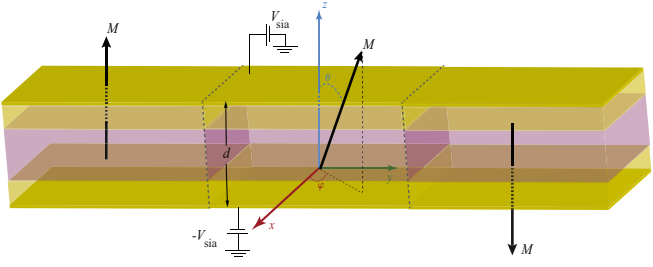}
\caption{Schematic view of a topological-insulator thin film's nanojunction consists of two domains with opposite magnetization polarized along the $z$-axis which is mediated by a central portion including magnetization with arbitrary polarized direction ${\bf M}$ as well as the structural inversion asymmetry ($V_{sia}$).}
\label{fig1}
\end{figure}

In this paper, to shed light on electronic transport in QAH insulators, we address a chiral selective tunneling through magnetic TI thin films with perpendicular ferromagnetic ordering which is responsible for some transport gaps in absence of SIA. It is well-known that in carbon nanotubes, conservation of the rotational symmetry of the incoming electron wave function\cite{rotational} and correspondingly, in even zigzag graphene nanoribbons\cite{cheraghchi_evenZGNR,cheraghchi_oddZGNR,parity,parity1,cheraghchi_nanojunction}, conservation of the transverse reflection symmetry of the incoming and outgoing wave functions result in some selection rules governing quantum transport. Here, quantum blocking of transport occurs when bands with opposite chiralities are aligned energetically by means of antiferromagnetic manipulation of electrodes. At zero SIA potential, the chirality attributed to each band is affected by out-of-plane magnetic field and also hybridization between states belonging to the upper and lower surfaces. Such a chiral selective tunneling in magnetic TI thin films fails by application of SIA potential or in-plane exchange field induced by ferromagnetically doped TI. To elucidate this selective tunneling, by means of Landauer-Buttiker formalism, transmission and also conductance is calculated for magnetic and electrostatic potential barriers and finally as an application of this chiral selective tunneling, a $p-n$ nanojunction switch is designed on magnetic TI thin film which shows a negative differential resistance in its $I-V$ curve.   

Organization of the paper is as the following: in Sec.\ref{sec2}, Hamiltonian and its discretized version along the transport axis is presented. Furthermore, based on this discretized Hamiltonian, Transmission and conductance is calculated by using non-equilibrium Greens function formalism. The band structure and chiral selective tunneling will be presented in Sec.\ref{sec3}. The current-voltage of a $p-n$ nanojunction in magnetic TI thin film is proposed in Sec.\ref{sec4} to show that there exists negative differential resistance in such systems. Finally we conclude in Sec.\ref{sec5}.

\section{Hamiltonian and Formalism}\label{sec2}
\subsection{Hamiltonian}
The effective low-energy Hamiltonian for the hybridized Dirac cones on the top and bottom surfaces of ultra-thin films of the magnetically doped (Bi,Sb)$_2$ Te$_3$ family materials near the $\Gamma$ point is described by\cite{effective}
\begin{equation}
\begin{aligned}
\centering
&\mathscr{H}(\textbf{k})=
\hbar v_f (k_y \sigma _x -k_x \sigma _y) \otimes \tau_z
+\Delta (\textbf{k}) \sigma _0 \otimes \tau _x+V_{sia}  \sigma _0 \otimes \tau _z
\\&+ \big ({\bf M. \sigma}\big )\otimes\tau _0
\end{aligned}
\label{eq:firsthamil}
\end{equation}
in the basis set of $|u,\chi_{+} \rangle ,|u,\chi_{-} \rangle ,|l,\chi_{+} \rangle ,|l,\chi_{-} \rangle$, where $u$, $l$ denote to the top and bottom surface states and $\chi_{+}$, $\chi_{-}$ stand for up and down spin states. In addition, $\tau_i$ and $\sigma_i$ ,$ (i = x, y, y)$ refer to the Pauli matrices in the surfaces and electron spin space respectively while $\sigma _0$ and $\tau_0$ are identity matrices. The mass term which is extracted by using the experimental data, is mapped on the form of function $\Delta(\textbf{k})=\Delta_0+\Delta_1 k^2$ for those TIs thinner than $ d=5 nm$ for Bi$_2$ Se$_3$ and [(Bi,Sb)$_2$ Te$_3$] family\cite{nature584,prb81041307}. The parameters $\Delta_0$ and $\Delta_1 $ are the fitted parameters of the band gap which depend on the thickness of TI thin film. Here, $\textbf{k}= (k_x , k_y )$, $k = |k|$ is the wave vector where its x-component is conserved during electron transport through a barrier which is constructed along the $y$-axis. $v_f$ shows the Fermi velocity which is considering to be structural symmetric on the top and bottom surfaces. The third term represents the structural inversion asymmetry (SIA) $V_{sia}$ which can be originated to the perpendicular electrical field applied on the surfaces and also to the potential difference induced by substrate\cite{nature584}. This SIA potential leads to the Rashba-like splitting of the energy spectra which introduces TI thin films as a good candidate for spintronic devices controlling by an electric field. The last term refers to the ferromagnetic exchange field which comes from the finite magnetization induced by magnetic impurities such as Ti, V, Cr and Fe which are doped in Bi$_2$Se$_3$ and [(Bi,Sb)$_2$ Te$_3$] family\cite{Magnetic}. The Landauer formalism presented in this paper is written in a general form such that conductance can be calculated when the induced ferromagnetism in TI thin films is directed along an arbitrary direction.

To investigate the chiral selective tunneling, a nanojunction consist of two domains with opposite magnetization polarized along the $z$-axis is considered such that it is mediated by a region including magnetization with arbitrary polarization ${\bf M}=(M_x,M_y,M_z)=M(\sin \theta \cos \phi, \sin \theta \sin \phi, \cos \theta)$ as well as the structural inversion asymmetry. A schematic view of TI thin film nanojunction under application of perpendicular applied bias is depicted in Fig.\ref{fig1}. Band spectrum arising from Hamiltonian \ref{eq:firsthamil} for those portions which have only $z$-axis magnetization $\theta=0$ is written as:
\begin{equation}
\begin{aligned}
E(k)=& \pm \Bigg((\hbar v_f k)^2+(\Delta(k))^2+M^2+V_{sia}^2 \\
& \pm 2 \Big[ (\hbar v_f k)^2 V_{sia}^2+M^2 \left(\left(\Delta(k)\right){}^2+V_{sia}^2\right) \Big] ^{\frac{1}{2}}\Bigg)^{\frac{1}{2}}
\end{aligned}
\label{eq:totalE} 				
\end{equation}

Note that Hamiltonian \ref{eq:firsthamil} which was written in the spin-surface Hilbert space can be rearranged in terms of the spin-orbital new basis set as $|\psi_b,\chi_{+} \rangle ,|\psi_{ab},\chi_{-} \rangle ,|\psi_b,\chi_{-} \rangle ,|\psi_{ab},\chi_{+} \rangle$ under a unitary transformation, which $\psi_b(\psi_{ab})$ denotes to the bonding (anti-bonding) orbital state originating from the hybridization between the top and bottom surface states.
\begin{equation}
\centering
H (\textbf{k})=\left(\begin{array}{cc}  h_{+} (\textbf{k}) & V_{sia}\sigma_x+M \sin \theta \;e^{i \varphi} \sigma_z \\ V_{sia} \sigma_x+M \sin \theta \;e^{-i \varphi} \sigma_z & h_{-}(\textbf{k}) \\\end{array} \right)
\label{eq:hamiltotalchiral} 				
\end{equation}
where the parity eigenstates have the following general form \cite{electrically}
\begin{equation}
\centering
\begin{aligned}
\vert \psi_b,\chi_{\pm}\rangle=&\left( \vert u,\chi_{\pm}\rangle+\vert l,\chi_{\pm}\rangle \right)/ \sqrt{2}\\
\vert \psi_{ab},\chi_{\pm}\rangle=&\left( \vert u,\chi_{\pm}\rangle-\vert l,\chi_{\pm}\rangle \right)/\sqrt{2}
\end{aligned}
\label{eq:3} 				
\end{equation}

In the absence of SIA potential ($V_{sia}=0$) and while $z$-axis magnetization (with $M_z=\pm M$ for $\theta=0,\pi$ conditions) is applied, the above effective Hamiltonian (Eq.\ref{eq:hamiltotalchiral}) reduces to two decoupled blocked diagonal matrix with opposite chirality\cite{PRL2013_zhang} as the following
\begin{equation}
\centering
h_{\alpha }(\textbf{k})=\text{$\hbar $v}_f \left(k_y \sigma _x- \alpha  k_x \sigma _y\right)+\Big(\Delta(\textbf{k})+\alpha  M_z \Big)\sigma _z
\label{eq:hamilchiral} 				
\end{equation}
where, $\alpha$ is the chiral index\cite{universality,geometrical,PRL2013_zhang} which has two different signs $\pm$ corresponding to the upper and lower block digonal matrices in Eq.\ref{eq:hamiltotalchiral}. Here the last term is the mass term which its negative sign leads to topologically non-trivial band structure. By diagonalization of the Hamiltonian \ref{eq:hamilchiral}, the following band dispersions would be obtained as
\begin{equation}
\centering
E_{\alpha }(k)=\pm \sqrt{\left(\text{$\hbar $v}_fk \right){}^2+\left(\text{$\Delta_0 $}+\text{$\Delta_1 $} k^2+\alpha M_z\right)^2}
\label{eq:Echiral} 				
\end{equation}
The energy gap for each band is $2|\Delta_0+\alpha M_z|$. The bands with positive mass term are trivial with a zero chern number if $|M_z|<|\Delta_0|$ and the bands with negative mass term are non-trivial with the chern number to be as $M_z/|M_z|$ if $|M_z|>|\Delta_0|$\cite{universality}.

Let us note that in the case of non-magnetic TI thin film ($M=0$) at zero SIA potential, $h_{-}(k)$ is the time reversed copy of $h_{+}(k)$ such that $h_{+}(k)=h^{*}_{-}(-k)$\cite{chiral_TI}. Although as it is well-known and seen of Eq.\ref{eq:hamilchiral}, the time-reversal symmetry breaks when any magnetization is turned on, system remains chiral-polarized even if out-of-plane magnetization is applied\cite{geometrical}. In fact, by application of $z$-axis magnetization, as shown in Fig.\ref{fig2}, the degeneracy of the bands with opposite chiralities breaks\cite{Phys_scri_2015}. 
\subsection{Landauer Formalism for Transport}
In order to study transport properties of TI thin films, one can numerically discretize Hamiltonian in the spin-orbital basis set $H (\textbf{k})$ (Eq.\ref{eq:hamiltotalchiral}) in a quasi one-dimensional lattice model with the new basis set $|k_x> \otimes \;|y_i> $, where $k_x$ is the $x$-component of the wave-vector\cite{graphene_dolfus}. In the tight-binding representation, the Hamiltonian in real space can be written as
\begin{equation}
\centering
\mathcal{H}=\sum_i [ \mathcal  {H}_{i,i}c_i^{\dagger}c_i+(\mathcal{H}_{i,i+\delta y} c^{\dagger}_i  c_{i+\delta y} +h.c.)]
\label{eq:7} 				
\end{equation}
where the diagonal term in discretized Hamiltonian is
\begin{equation}
\begin{aligned}
\centering
&\mathcal{H}_{i,i}=-t_a E_x \sigma _y +\bar{m}\;\sigma _z \otimes \beta_0+V_{sia} \sigma_x\otimes \beta_x +\sigma _z \otimes \big ({\bf M. \beta}\big )\label{eq:8}
\end{aligned}				
\end{equation}
 and their hopping terms are derived as the following
\begin{equation}
\centering
\mathcal{H}_{i,i+\text{$\delta $y}}=(-B_a \sigma _z+i t_a \sigma _x)\otimes \beta_0.
\label{eq:9} 				
\end{equation}
Here $c^{\dagger}_i$ ($c_i$) is the creation(annihilation) operator of electron at the site index $i$ and $\text{$\delta $y}$ represents unit vectors along the $y$ directions. The lattice constant along the $y$-axis is $a=y_\text{$i+1$} -y_i$ which is assumed to be $1 nm$ for all numerical calculations. The Pauli matrices $\beta_i$, ${i=x,y,z}$, which are different of the spin or surface Pauli matrices, are defined for the spin-orbital Hilbert space and $\beta_0$ refers to the identity matrix. Here parameters $E_x$, $t_a$, $B_a$ and $\bar{m}$ are in relation with the parameters in Eq.(\ref{eq:firsthamil}) as $E_x={2a}{k_x}$, $t_a=\hbar v_f/{2a}$, $B_a={\Delta_1 }/{a^2}$ and $\bar{m}=\frac{1}{4} B_a E_x^2+2 B_a+\Delta _0$.
To calculate transport properties, we applied the two terminal Landauer formula by using non-equilibrium Green function formalism at zero temperature\cite{buttiker}. Transmission is calculated numerically by
\begin{equation}
\centering
T(E,k_x)=\text{Tr}\left[\Gamma _L G^r  \Gamma _R G^a\right]
\label{eq:10} 				
\end{equation}
where broadening of the energy states arising from the attached electrodes is $\Gamma_{R(L)}=i \left(\Sigma ^r _{R(L)}-\Sigma ^a_{R(L)}\right)$, and $\Sigma ^{r(a)}_p$ is the  retarded/advanced self-energy describing the coupling between the right/left $p^{th}$ semi-infinite lead (p = L, R) and the central region. We numerically evaluate the lead self-energies using the iterative technique developed by Lopez Sancho {\it et. al.}\cite{lopez}. $G^r=\left(G^a\right)^+$ is the retarded Green’s function and can be calculated by using the formula $G^r(E)=\Big((E+i \eta) {\bf 1}-\mathcal{H}_c-\Sigma^r _L-\Sigma ^r_R\Big)^{-1}$, where $\eta$ is $0^{+}$. Here $\mathcal{H}_c$ is the Hamiltonian for the central scattering region and ${\bf 1}$ is the identity matrix.

Conductance is calculated by the angularly averaged transmission which is projected along the current flow direction,
\begin{equation}
\centering
G(E_f)/G_0=\int_{-\pi/2}^{\pi/2} T(E_f,\eta) \cos \eta d \eta
\end{equation}
where $\eta$ is the incident angle of electrons into the barrier which is defined as $ \eta=\tan^{-1} (k_x/k_y)$. $G_0$ is $N e^2/\hslash$ where $N$ is the channel Number. To check transmission calculations, all results reported in Ref.\cite{transportTI_nanoscale} were derived as a special case of our general model.
\section{Chiral Selective Tunneling}\label{sec3}
To present the chiral selection rule which governs transport properties of magnetically doped TI thin films, at first, let us calculate transmission in a special case of zero SIA, $V_{sia}=0$ and also assuming magnetization to be directed along the $z$-axis. In this case, Hamiltonian is block diagonal in the spin-orbital basis set as presented in Eq.\ref{eq:hamiltotalchiral} and the bands are decoupled from each other with opposite chirality $\alpha$. To describe the experimental spectrum, the fitted parameters are taken of the values sorted in the table \ref{tab:parameter}. In this study, four quintuple layers of $(Bi_{0.1} Sb_{0.9})_2 Te_3$ ($\Delta_0=-0.029 eV,\Delta_1=12.9 eVA^2$) with $\Delta_0 \Delta_1<0$ is considered. However, it should be noted that our proposal for the chiral selective tunneling is still valid for the case of $\Delta_0 \Delta_1>0$.

\begin{table}
\begin{center}
\begin{tabular}{|c | c | c | c | c|}
\hline
&      $\hbar v_f(eVA^{0})$&	$\Delta_0(eV)$&	$\Delta_1(eVA^2)$ \\ \hline
4QLs&	 2.36&-0.029&12.9\\
3QLs&	3.07 &+0.044&37.3\\
\hline
\end{tabular}
\end{center}
\caption{The parameters of the 2D effective Hamiltonian in
Eq.(1) for low energy physics in $(Bi_{0.1} Sb_{0.9})_2 Te_3$ thin films
with different thickness.\cite{electrically}}\label{tab:parameter}
\end{table}

\begin{figure}[!ht]
\centering
\includegraphics[width=1.0\linewidth]{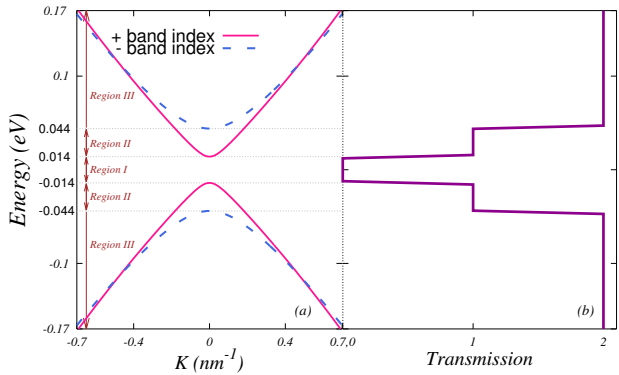}
\caption{a) Band structure of magnetically doped-TI thin film when structural inversion asymmetry is zero. Here the exchange field induced by magnetic impurities is assumed to be $M=0.015eV$ and $\theta=0$. b) Transmission coefficient in terms of Fermi energy.}
\label{fig2}
\end{figure}

At the first stage, no barrier or step function is considered along the current flow direction. The band spectrum in this case is plotted in Fig.\ref{fig2}{\bf a} in which each band is marked by its chiral index ($\pm$) deriving in Eq.\ref{eq:Echiral}. The inner/outer bands belong to the $+(-)$ index which is depicted by pink/blue solid/dashed lines in the figure. The edges of the bands with the chiral index $\alpha$ occur at energies $\pm |\Delta_0+\alpha M_z|$. Since both parameters $\Delta_0$ and $M_z$ could be experimentally positive or negative values\cite{universality}, one can summarize determination of the chiral index attributed to each band based on the following expression: {\it the chiral index of the inner bands is in the opposite sign with the sign of $\Delta_0 M_z$, while the chiral index of the outer bands corresponds to the sign of $\Delta_0 M_z$.} As an example, the index of the inner bands in Fig.\ref{fig2}a is $\alpha=+1$ because $\Delta_0 M_z<0$.

For the case of $\Delta_0 \Delta_1<0$, the band branches with opposite chiralities are crossed each other at a finite $k=\sqrt{|\Delta_0/2\Delta_1|}$\cite{prl2013QT} as seen in Fig.\ref{fig2}a.

\begin{figure}[!ht]
\includegraphics[width=1.0\linewidth]{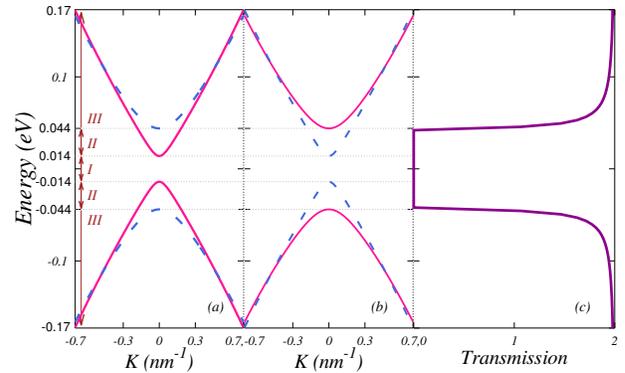}
\caption{Band structure of magnetically doped-TI thin film when structural inversion asymmetry is zero. Here the exchange field induced by magnetic impurities is assumed to be $M=0.015eV$ with a) $\theta=0$ for the left side of the step function and b) $\theta=\pi$ for the right side of the step function.}
\label{fig3}
\end{figure}

\begin{figure}[!ht]
\includegraphics[width=0.47\linewidth]{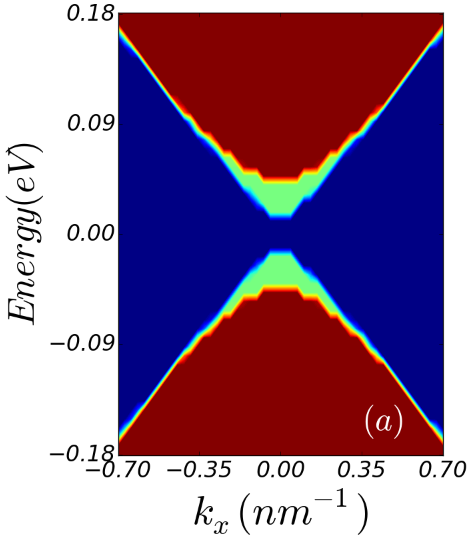}
\includegraphics[width=0.445\linewidth]{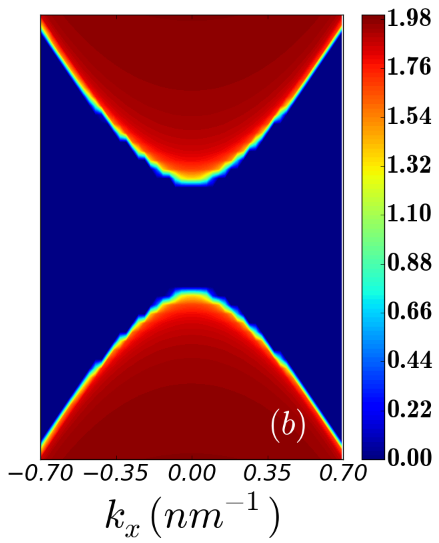}
\caption{3D contour plot of transmission in terms of Fermi energy and incident angle of electrons ($k_x$) for TI thin film described in a) Fig.\ref{fig2} and b) Fig.\ref{fig3}.}
\label{fig3d}
\end{figure}

 Let us first consider normal incident electrons hitting into the step function with $k_x=0$, although our expression about selective tunneling can be extended to the case of $k_x \ne 0$. Depending on the channel number in each Fermi energy, transmission through such magnetic TI thin film is divided into three regions in Fig.\ref{fig2}a. The transport gap originated from the band gap occurs around the band center in a range [$-|\Delta_0+M_z|$,$|\Delta_0+M_z|$]. This gap region is called the region type $I$. Besides, there are ranges [$|\Delta_0+M_z|$,$|\Delta_0-M_z|$] and [$-|\Delta_0-M_z|$,$-|\Delta_0+M_z|$] with transmission coefficient $T=1$ which corresponds to one transport channel with (${\bf +}$) chiral index. This range is called the region type $II$. Finally transmission reaches to its maximum value $T=2$ for the energy region $III$ in which there are two allowed transport channels. To have complete view, a 3D contour-plot of transmission in terms of Fermi energy and the incident angle of electrons ($k_x$) are depicted in Fig.\ref{fig3d}a. This graph shows that transmission follows the band spectrum. Moreover, in this case, it was checked that conductance as a function of Fermi energy has the same behavior as transmission.
\\
Now, to demonstrate the forbidden electron transition through the bands with opposite chirality, let us design a step function of
magnetic potentials as depicted in Fig.\ref{fig1} without the central portion such that at the region $y<0$, the magnetization vector is $(0,0,M)$ with $\theta=0$ and for the region $y>0$, we have $(0,0,-M)$ with $\theta=\pi$ meaning that the magnetization direction is becoming inverted along the step function. This band engineering causes to align energy states with opposite chiralities belonging to different sides of the step function. 
\\
Fig.\ref{fig3} shows the band structure of the left and right sides of the magnetic step, as well as transmission coefficient through this magnetic step in terms of Fermi energy. In this turn, transmission gap is increased to $2(|\Delta_0|+|M_z|)$ while still there exist band states inside the energy region $II$. In the region $III$, transmission reaches to its maximum value $T=2$ according to two transport channels. The 3D contour-plot of transmission in terms of Fermi energy and $k_x$ in Fig.\ref{fig3d}b reveals us that the forbidden electron transition between two band states with opposite chiralities would be preserved at other incident angles $k_x \neq 0$, too.

{\it Transition Rule:} Focusing on the region $II$ shows that the transition probability ($t$) from the energy state with ($-$) chiral index $\psi_-=(\psi_b\chi_{+},\psi_{ab}\chi_{-})^\dagger$ to the state with the opposite chiral index $\psi_+= (\psi_b\chi_{-},\psi_{ab}\chi_{+})\dagger$ through the $z$-axis magnetic step potential can be written as
\begin{equation}
\centering
t=\left\langle \Psi_L|(2M \sigma_z \otimes \beta_z) | \Psi_R \right\rangle=0
\label{eq:11} 				
\end{equation}
where the state belonging to the left electrode is as $\Psi_L=(\psi_+,0)^{\dagger}$ and the right state is $\Psi_R=(0,\psi_-)^{\dagger}$. Therefore, transition probability is apparently zero. So transmission ($T\propto|t|^2$) is blocked in the energy regions $II$. However, we will show later that these forbidden transport channels are opened if an in-plane magnetic step potential is applied on TI thin films. 
\begin{figure}[!ht]
\includegraphics[width=1.0\linewidth]{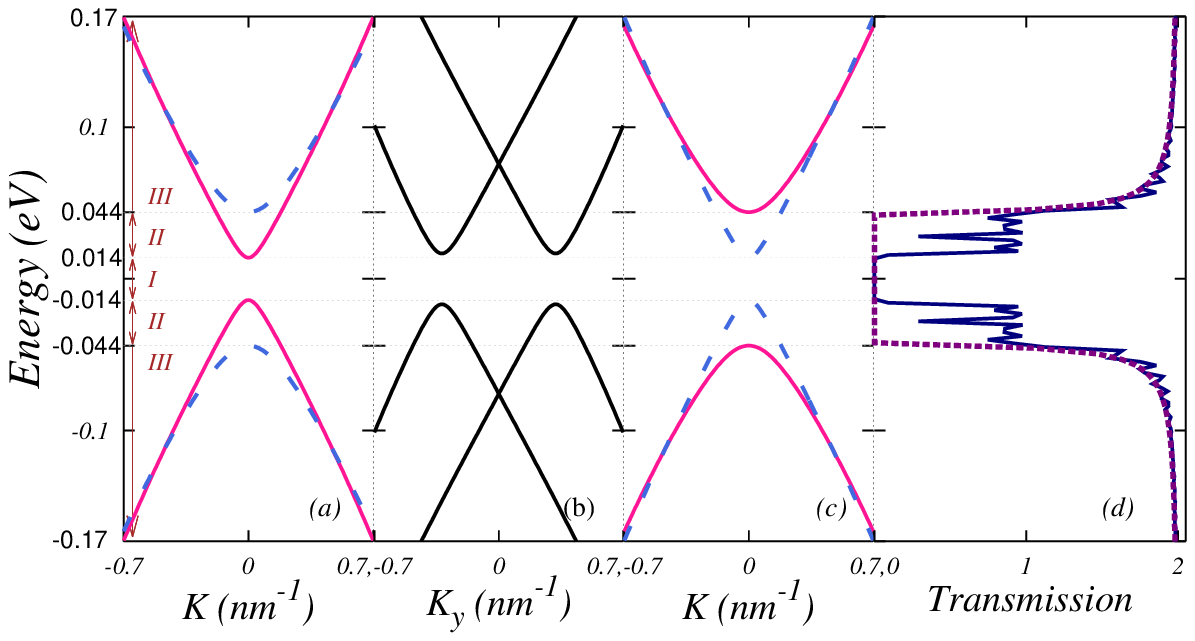}
\includegraphics[width=1.0\linewidth]{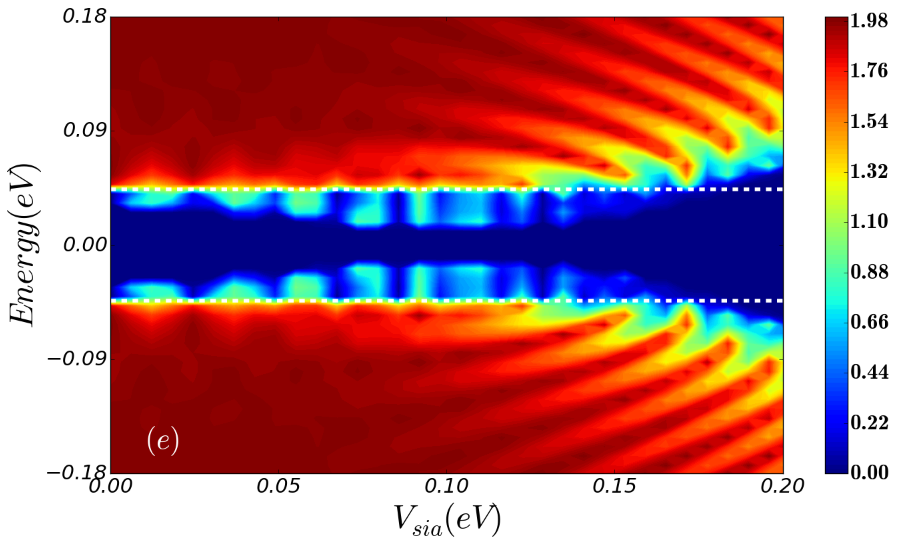}
\caption{Band structure of a nanojunction designed on magnetically doped-TI thin film when structural inversion asymmetry is applied on the central portion sandwiched in between two electrodes with opposite $z$-axis magnetization. Here the central portion (b) is assumed to be non-magnet $M=0$ but with non-zero structural inversion asymmetry $V_{sia}=0.07eV$. The exchange field induced by magnetic impurities inside the electrodes is assumed to be $M=0.015eV$ where for the left electrode $\theta_L=0$ c) and for the right electrode $\theta_R=\pi$. d) Transmission coefficient in terms of Fermi energy is plotted for $k_x=0$. For comparison, transmission for the magnetic step of Fig.\ref{fig3} is also plotted by dashed lines. The length of the central portion is considered to be $80 nm$. e) 3D contour-plot of transmission in terms of Fermi energy and structural inversion asymmetry at $k_x=0$, the area between two white dashed lines belongs the transport gap.}
\label{fig4}
\end{figure}
\begin{figure}[!ht]
\includegraphics[width=1.0\linewidth]{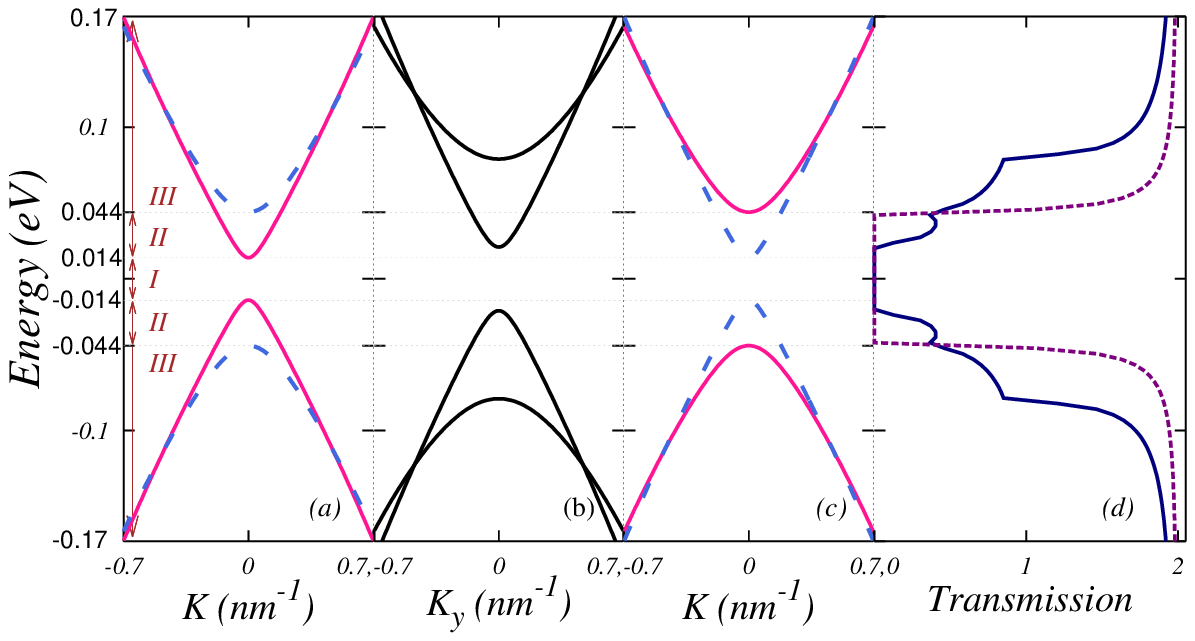}
\includegraphics[width=1.0\linewidth]{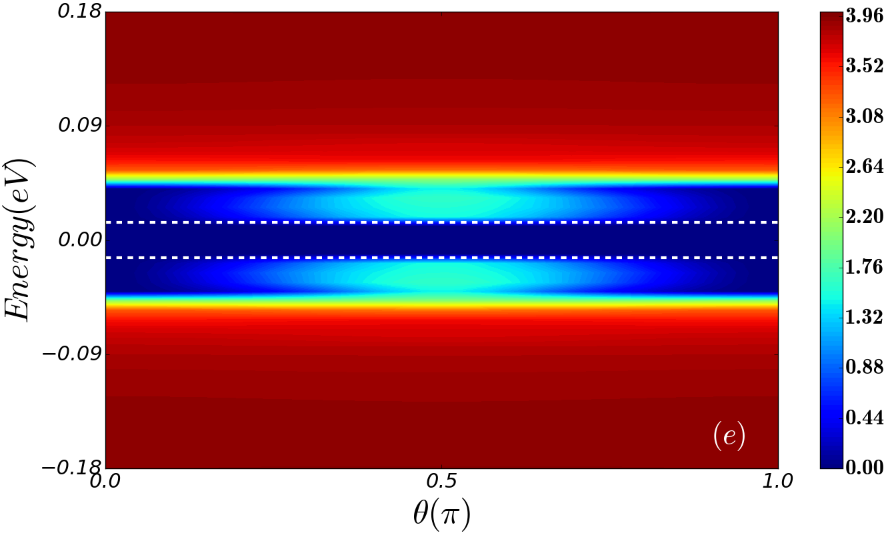}
\includegraphics[width=1.0\linewidth]{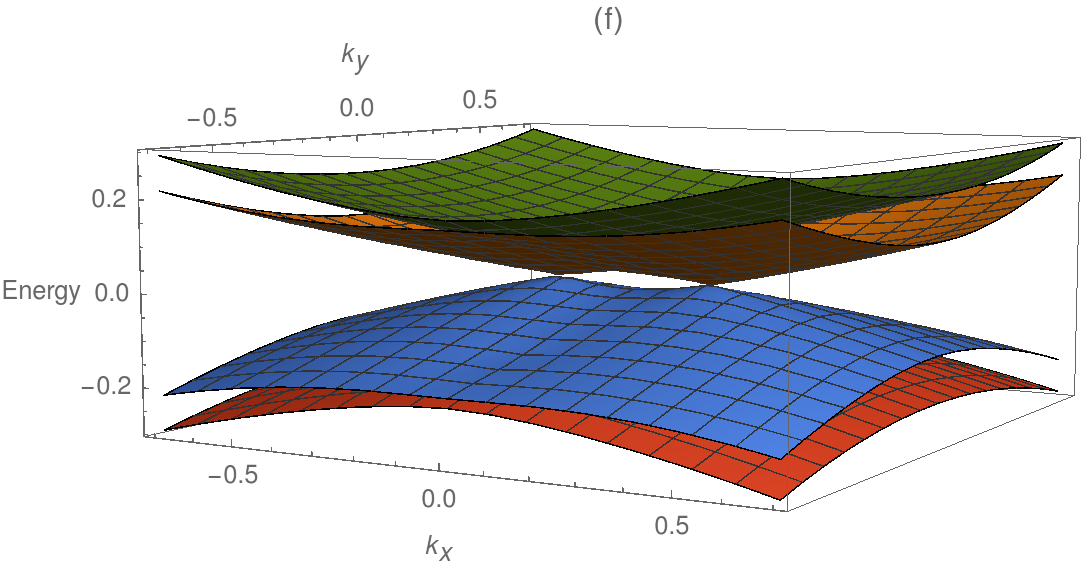}
\caption{Band structure of a nanojunction designed on magnetically doped-TI thin film when the $x$-axis magnetization ($M_x$) is applied on the central portion sandwiched in between two electrodes with opposite $z$-axis magnetization. Here in the central portion, $M_x=0.05eV$ with $\theta=\pi/2,\phi=0$ and for $V_{sia}=0$. The exchange field induced by magnetic impurities inside the electrodes is assumed to be $M=0.015eV$ where a) for the left electrode $\theta_L=0$ and c) for the right electrode $\theta_R=\pi$. d) Transmission coefficient in terms of Fermi energy is plotted for $k_x=0$. For comparison, transmission for the magnetic step of Fig.\ref{fig3} is also plotted by dashed lines. The length of the central portion is assumed to be $55 nm$.
 e) 3D contour-plot of conductance $G/G_0$ in terms of Fermi energy and polar angle $\theta$ at $k_x=0$ when magnetization is rotated in the $x-z$ coordinate plane ($\phi=0$). The area between two white dashed lines belongs to the gap in the electrodes band structre. f) 3D plot of band structure in terms of $k_x$ and $k_y$ for $M_x=0.05eV$.}
\label{fig5}
\end{figure}

To realize conservation of the chirality, one can define the following unitary operator in the spin-orbital basis set,
\begin{equation}
\centering
 C= \sigma_0\otimes \beta_z
\label{eq:12}
\end{equation}
where $C^2=1$. This operator in the spin-layer basis set is proposed to be as $\sigma_z \otimes \tau_x$ which commutes with Hamiltonian (Eq.\ref{eq:firsthamil}) at $V_{sia}=0$ and $\theta=0$. The same also occurs for Hamiltonian (Eq.\ref{eq:hamiltotalchiral}) which commutes with $C$ operator,
\begin{equation}
\centering
[ H (\textbf{k}) ,C ] =0
\label{eq:13}
\end{equation}
leading to a conservation rule in the chirality. The gate voltage also preserves this conservation rule. Therefore, transition rate between the energy states with opposite chiralities is blocked through an application of the gate voltage, $\left\langle \Psi_L|(V_g \sigma_0 \otimes \beta_0) | \Psi_R \right\rangle=0$. However, one can verify that in the presence of the SIA potential, Hamiltonian does not commute with $C$ operator which leads to the bands with mixture of different chiralities,
\begin{equation}
\centering
[  V_{sia} \sigma_x\otimes \beta_x  ,C ] \neq 0.
\label{eq:13}
\end{equation}
Correspondingly, in the presence of SIA, one can check that the transition rate from the left-side states to the right-side states with opposite bchiralities is not zero giving rise to break the conservation rule of $\alpha$.
$$\left\langle \Psi_L|(V_{sia} \sigma_x \otimes \beta_x) | \Psi_R \right\rangle \neq 0$$
{\it SIA Potential}: To demonstrate the above proposition, let us focus on the transmission coefficient shown in Fig.\ref{fig4}. A portion which is under  application of the SIA potential, is placed between two magnetic electrodes with opposite perpendicular polarizations. The band gap of the central portion is justified to be nearly equal to the electrode's energy gap. What is clear of Fig.\ref{fig4}d is that the SIA potential can change the chiral index ($\alpha$) of the state. In the region $II$ in which there exists a transport gap in Fig.\ref{fig3}, by turning the SIA potential on, transmission would be nonzero and equal to unity arising from one transport channel accessible in this region. To have a complete view, at $k_x=0$, 3D contour-plot of transmission in terms of Fermi energy and SIA potential is depicted in Fig.\ref{fig4}e to show opening of some transport channels inside the gap. The overall band gap is increased by the SIA potential after a critical SIA potential in which the band gap in the central portion exceeds the transport gap of electrodes. As it is shown in this figure, transmission oscillates with the SIA potential without decay which is originated from the resonance condition in the central portion. 

{\it In-plane Magnetization}: To end this section, let us look at the effect of in-plane magnetization on transition rule. As it is seen in Eq.\ref{eq:hamiltotalchiral}, in-plane magnetization leads to mixing of different chiralities $\alpha$ with each other. The in-plane magnetization along the $x$ and $y$ axis in the spin-orbital basis set are represented as $M_x=\sigma_z \otimes \beta_x$ and $M_y=\sigma_z \otimes \beta_y$, respectively. One can simply check that $[M_x,C] \neq 0$ and $[M_y,C] \neq 0$ which means that in-plane magnetization does not preserve $\alpha$. This fact can be confirmed by calculation of the transition rate between two states with opposite chiralities in the presence of an in-plane magnetization, $\left\langle \Psi_L|M_{x/y} | \Psi_R \right\rangle \neq 0$.
\\
A confirmation of such claims is displayed in Fig.\ref{fig5} in which transmission is plotted in terms of Fermi energy for a nanojunction consisting of a portion with an in-plane magnetization $M_x$ tilted with $\theta=\pi/2$ sandwiched between two electrodes with opposite $z$-axis magnetization ($\theta_L=0$ and $\theta_R=\pi$ for the left and right side electrodes). Again one can observe that in the region II of Fig.\ref{fig5}d where it is expected to have blocked transport channels, in-plane magnetization causes to induce non-zero transmission. Conductance which is defined as the incident angle-averaged transmission projected along the current flow direction is calculated in Fig.\ref{fig5}e in respect to the polar angle when magnetization is rotated in the $x-z$ coordinate plane. As long as the polar angle $\theta$ tends to $\pi/2$, the transport blocking induced by conservation of chirality is getting to be failed and conductance is enhancing. However, transport is blocked when magnetization tends to orient along the out of plane direction.   
\\
\begin{figure}[!ht]
\centering
\includegraphics[width=0.4\textwidth]{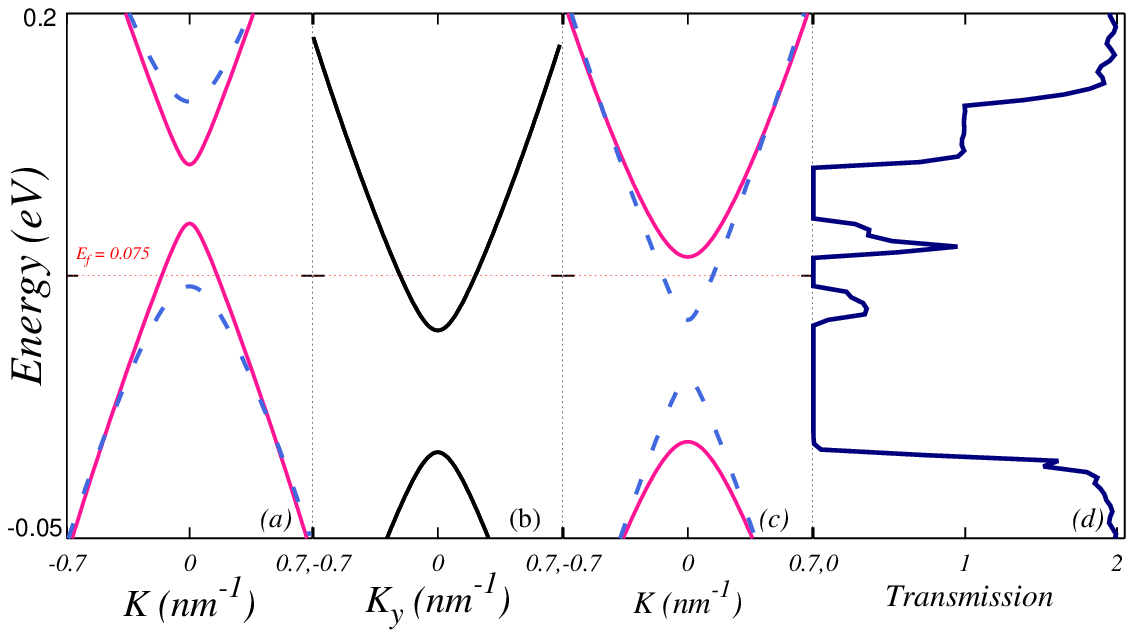}
\includegraphics[width=0.4\textwidth]{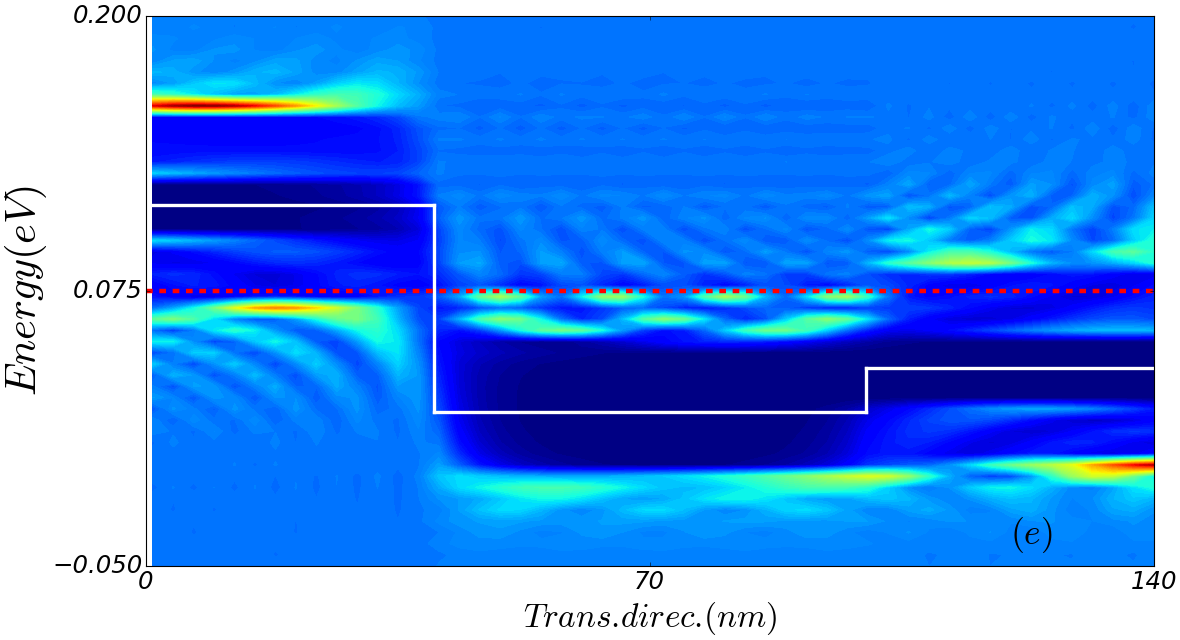}
\includegraphics[width=0.395\textwidth]{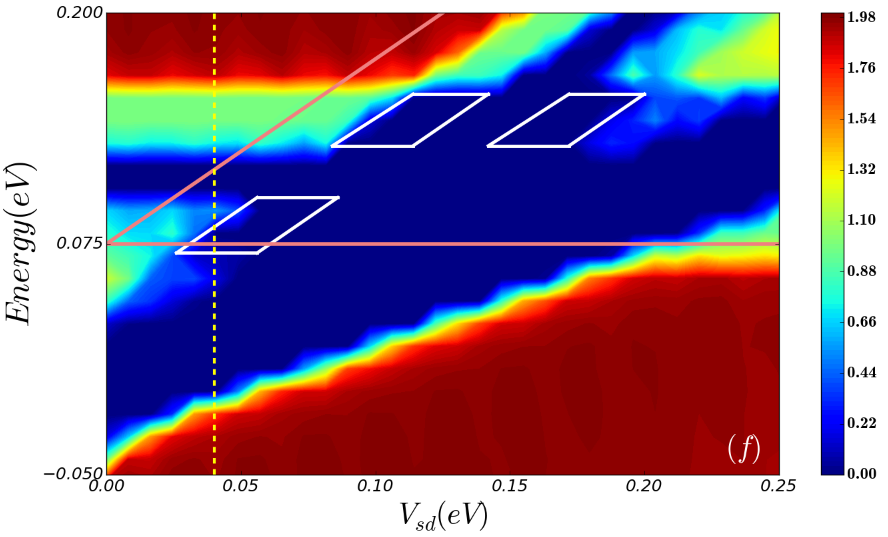}
\includegraphics[width=0.43\textwidth]{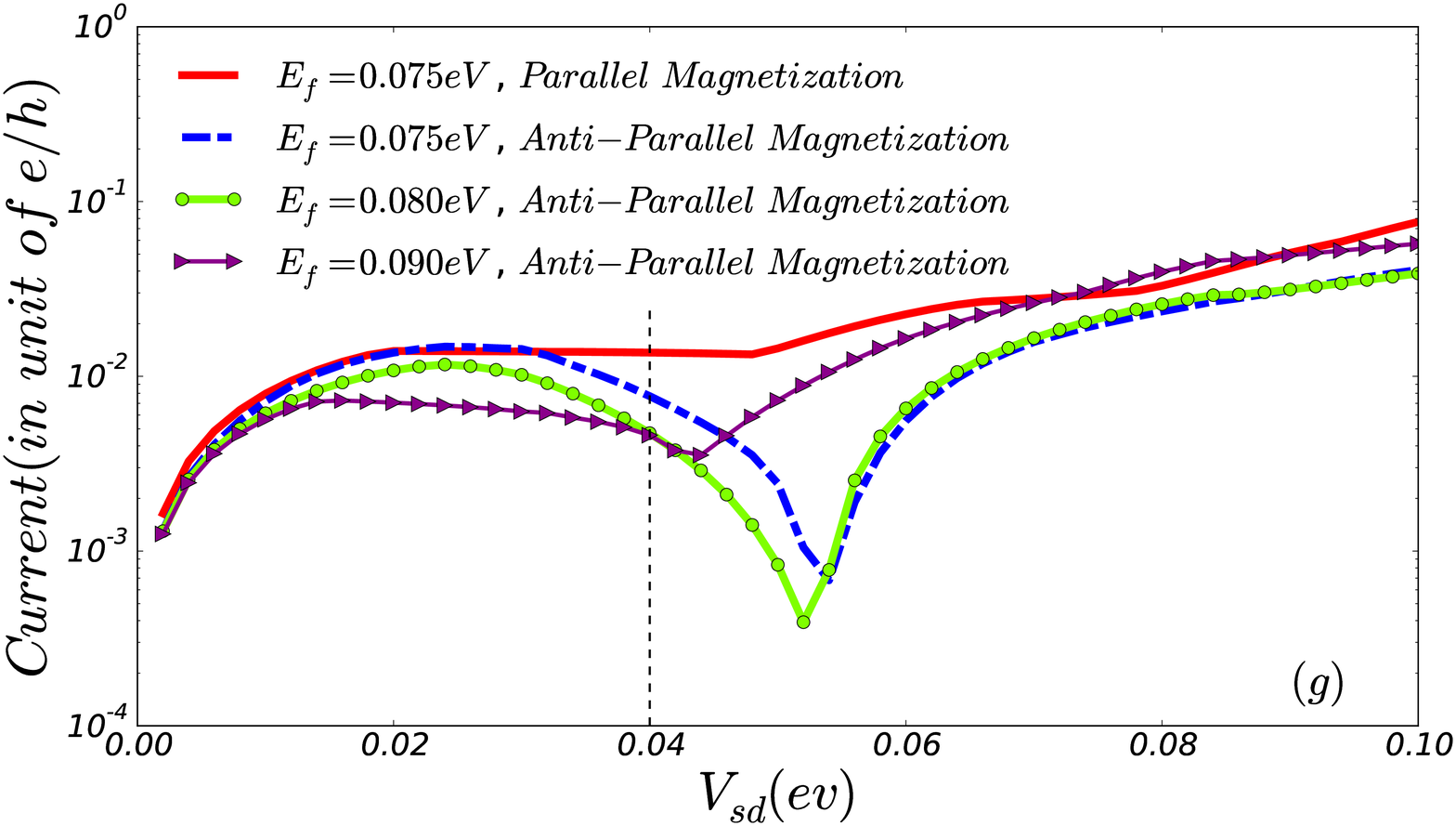}
\caption{a,b,c) The band structure of three regions of a $p-n$ nanojunction composed of two electrodes with opposite $z$-axis magnetic field mediated by a non-magnet portion. Here,  the normal incident is studied $k_x=0$.   d) Transmission coeffiecient in respect to energy at special source-drain bias $V_{sd}=0.04 eV$ shown as dashed vertical line in 3D plot of panel (f). e) Local density of states on each site for $V_{sd}=0.04 eV$. and all three regions.  f) 3D contour-plot of transmission as function of energy and source-drain bias $V_{sd}$. Pink solid lines show the energy integration window of transmission coefficient. Diamond marks refer to those blocked transport channels which are induced by the chiral selection rule. g) Current-voltage ($I-V_{sd}$) characteristic curve for different Fermi energies.}
\label{fig6}
\end{figure}

 \begin{figure}[!ht]
\centering
\includegraphics[width=0.5\textwidth]{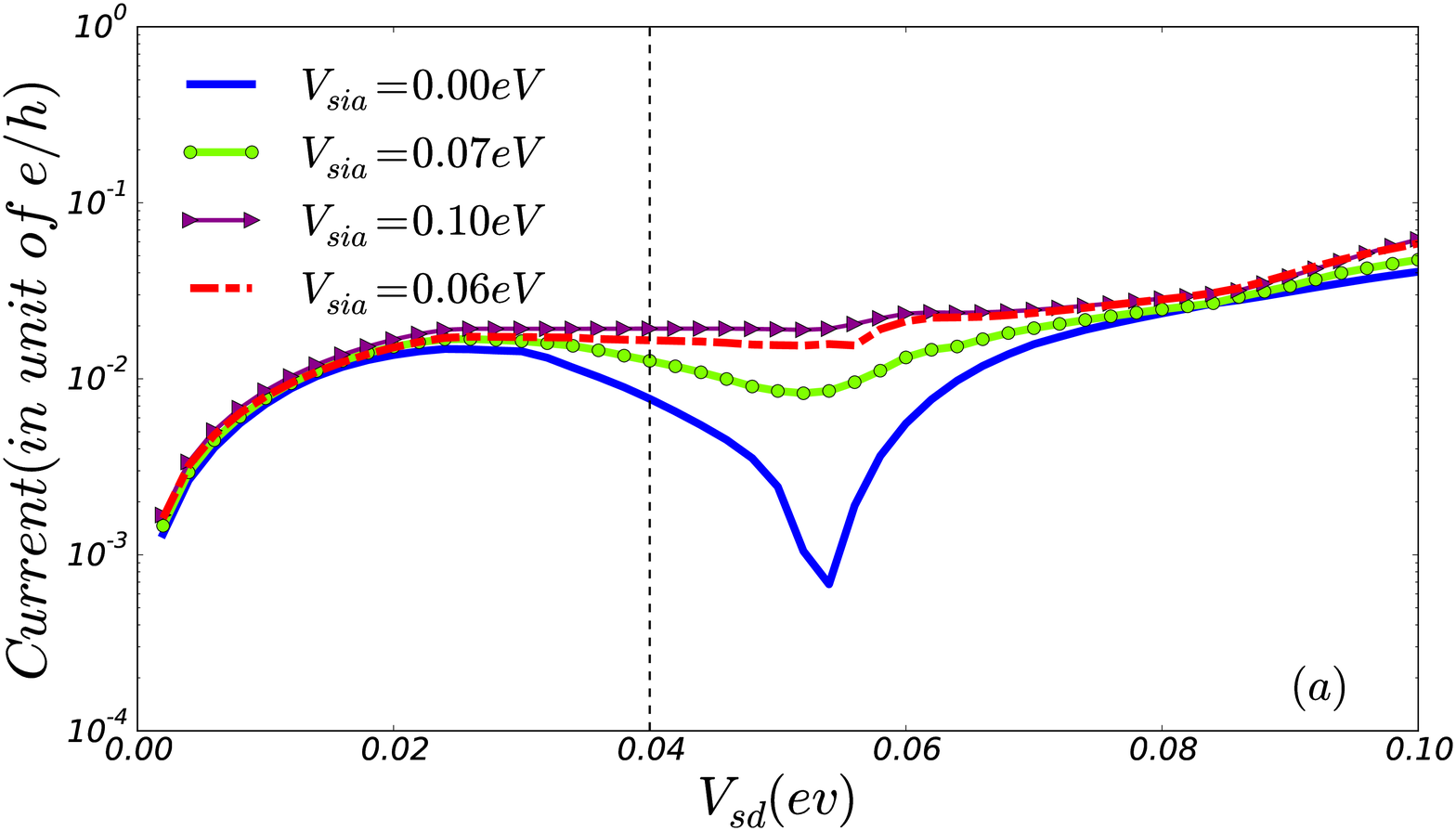}
\includegraphics[width=0.46\textwidth]{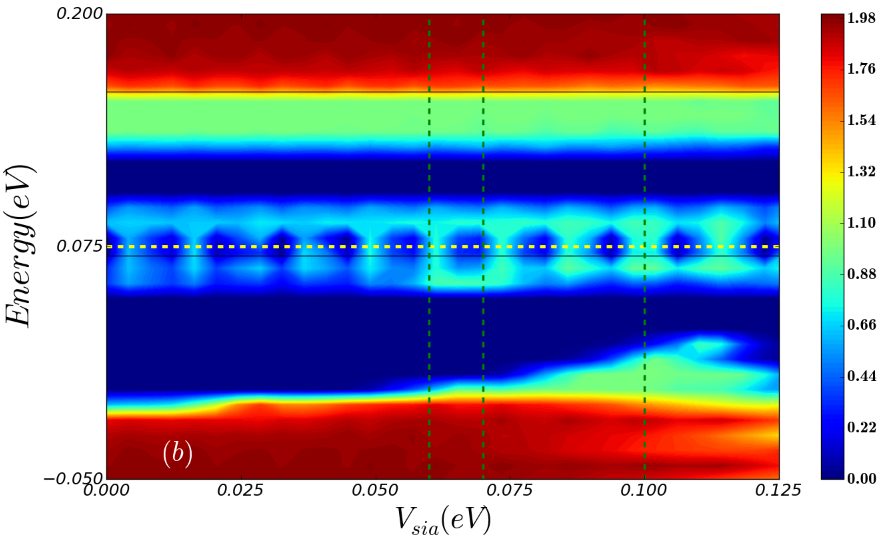}
\caption{a) Current-voltage characteristic curve ($I-V_{sd}$) for a $p-n$ nanojunction composed of two electrodes with opposite $z$-axis magnetic field mediated by a non-magnet portion under application of different structural inversion asymmetry (SIA) potentials. b) 3D contour-plot of transmission coefficient in respect to the energy and SIA potential at special source-drain bias $V_{sd}=0.04 eV$. Vertical lines correspond to three SIA potentials shown in the panel (a). Horizontal dashed line represents the Fermi level. }
\label{fig7}
\end{figure}

The magnetic band gap which is opened for out of plane magnetization, would be closed if magnetization is in-plane. However, at normal incident $k_x=0$, there exists a band gap which is depicted in the band structure of the central portion shown in Fig.\ref{fig5}b. In the presence of in-plane magnetization $M_x$ and $M_y$, the band spectrum is written as the following:
\begin{equation}
\begin{aligned}
\centering
&E(k)=\pm\Bigg((\hbar v_f k)^2 +M^2+\left(\Delta _0+\Delta _1 k^2\right){}^2
\\& \pm2\Big[(\hbar v_f k_x)^2(M_x^2+M_y^2)+\left(\Delta _0+\Delta _1 k^2\right){}^2(M^2)\Big]^\frac{1}{2}\Bigg)^\frac{1}{2}			
\end{aligned}
\label{eq:Einplan}
\end{equation}
where $M^2=M_x^2+M_y^2+M_z^2$. The edges of the conduction and valence bands at $k_x=0$ occur at $\pm ||\Delta_0|-M|$. The band structure as function of $k_x$ and $k_y$ which is presented in Fig.\ref{fig5}f gives us a whole view of the spectrum for topological-insulator thin films with a magnetization directed along the $x$-axis. The closing of the band gap is clear, however, it is gapful at normal incidence $k_x=0$.  
\section{Negative Differential Resistance}\label{sec4}
Recently, a proposal for possible manipulation of transistor devices on TI thin films was designed and presented by Ref.\cite{electrically} which is based on the metallic dissipation-less edge states appearing in the QAH phase. Indeed, the SIA potential can induce a phase transition from the QAH to NI phase which makes feasible having on and off current. The on-current arises from the chiral edge states while the off-current is originated to the band gap in NI. However, in this work, a nano-switch is engineered by means of magnetic TI thin films in which its off-current is dominantly caused by the forbidden transition between quantum states with opposite chiralities. 

Let us consider at first a $p-n$ nanojunction which is composed of two electrodes with opposite $z$-axis magnetization mediated by non-magnet portion where there is no Rashba splitting of the spectrum arising from the SIA potential. Although, based on the band gap emerging in electrodes, one can design an Esaki-like diode\cite{Esaki} to search for nano-electronic switch\cite{transportTI_nanoscale}, we try to seek for negative differential resistance (NDR) induced by the chiral selective tunneling. The left electrode is gated to the higher voltage as $V_g=0.114 eV$ in compared to the right electrode. The Fermi energy is set to be as $E_f=0.075 eV$ which leads to $p$-type semiconductor in the left and $n$-type semiconductor in the right electrode. Since the SIA potential is absent in the middle portion, there is no Rashba splitting in the spectrum as shown in Fig.\ref{fig6}b while the left and right electrodes are energetically gate-shifted against each other to seek for an energy alignment between two states with opposite chiralities. A transport gap at the Fermi energy is seen in Fig.\ref{fig6}d for the case of $k_x=0$. This blocked channel would be openned if the magnetization in electrodes is parallel. A source-drain bias $V_{sd}$ is applied along the system such that the bias voltage applied on the non-magnet and the right electrode portions is supposed to distribut as $V_{sd}/2$ and $V_{sd}$, respectively. The local density of state of this structure is plotted in Fig.\ref{fig6}e in which white solid lines is drawn for the eyes guide of the potential profile. The valley around the edge of the potential barriers refers to the band gaps of each region. However, thanks to the the chiral selective tunneling, around the Fermi level where a transport gap emerges, resonant states with opposite chiralities in electrodes disappear while there exist significant resonant states in the non-magnet portion (the middle one). This region corresponds to the region $II$ shown in Fig.\ref{fig2}.

The transport properties of such $p-n$ junction is affected by $V_{sd}$. A 3D contour-plot of transmission in respect to the energy and $V_{sd}$ is represented in Fig.\ref{fig6}f. In this figure, in addition to the three band gaps (blue regions) belonging to the three portions, there exist some regions marked by diamond where in these regions, transmission is blocked due to the transition rule. By increasing the source-drain bias, the left electrode's gap is fixed while the middle and right side gaps are shifted linearly in energy with $V_{sd}/2$ and $V_{sd}$, respectively. We have also checked that these diamonds-like regions are filled with the opened transport channels for the case of parallel magnetization in electrodes. 

The current-voltage characteristic curve is presented in Fig.\ref{fig6}g for different values of Fermi energy. Negative differential resistance occurs at a voltage corresponding to the region marked by diamond in the panel (g) of Fig.\ref{fig6}. It means that this NDR is originated to conservation of the chiral index. By increasing the Fermi energy, NDR disappears. Indeed, an increase in Fermi energy causes to fall the blocked transport region (induced by the conservation of chirality $\alpha$) out of the integration window $[E_F,E_F+V_{sd}]$. It was also checked that NDR disappears if magnetization in electrodes is parallel with each other.

By application of the structure inversion asymmetry on the middle part of this system, however, it is expected to fail the chiral selective tunneling giving rise to open the blocked transport channels shown in the diamond-shape marks of Fig.\ref{fig6}f. In this case, as we explained in Fig.\ref{fig6}, the band structure is the Rashba-splitting while the system is not chiral-polarized and contains a mixture of chiralities. The current-voltage characteristics curve presented in Fig.\ref{fig7}a demonstrates that NDR is getting weak as long as the SIA potential is increasing. However, this is not the whole picture, because by looking at the 3D-plot of transmission coefficient in terms of energy and SIA potential (Fig.\ref{fig7}b), it is seen that transmission at the Fermi energy has an oscillating behavior with the SIA potential without decaying arising from topological surface states\cite{transportTI_nanoscale}. This oscillation is related to the resonant states which gives rise to constructive and destructive interferences depending on the SIA potential.           
 
\section{Conclusion}\label{sec5}
The quantum transport properties through a thin films of magnetically doped topological-insulator is investigated by using Landauer formalism. The main message of this work is addressing a chiral selective tunneling appearing when out-of-plane magnetization of neighbouring regions is directed in opposite. In this case, transition between band states with opposite chiraities is blocked. These quantum transport gaps induced by conservation of chirality will be as opened channels if an in-plane magnetization is turning on. Moreover, this transition rule fails when a structural inversion asymmetry is present. Based on this selective tunneling one can design a $p-n$ nanojunction in which transport gaps result in negative differential resistance emerging in the current-voltage characteristics curve. By applying the SIA potential on $p-n$ nanojunction, it is demonstrated that this NDR is getting weak.

\section*{Acknowledgment }
H.C. thanks the International Center for Theoretical Physics (ICTP) for their hospitality and support in capacity of associate member of the center.


\begin{thebibliography}{99}
\bibitem{prl95Q} 
C. L. Kane and E. J. Mele, Phys. Rev. Lett. {\bf 95}, 226801 (2005).
\bibitem{prl96} 
B. A. Bernevig and  S. C. Zhang, Phys. Rev. Lett. {\bf 96}, 106802 (2006).
\bibitem{prl95Z} 
C. L. Kane and E. J. Mele, Phys. Rev. Lett. {\bf 95}, 146802 (2005).
\bibitem{prl97} 
S. Murakami, Phys. Rev. Lett. {\bf 97}, 236805 (2006).
\bibitem{science314} 
B. A. Bernevig, T. L. Hughes and S. C. Zhang, Science {\bf 314}, 1757–1761 (2006).
\bibitem{science318} 
M. König, {\it et. al.} Science {\bf 318}, 766–770 (2007).
\bibitem{nature398} 
Y. Xia, {\it et. al.} Nat. Phys. {\bf 5}, 398 (2009).
\bibitem{science329} 
R. Yu, Wei Zhang, H-J. Zhang, S-C. Zhang,  X. Dai, Z. Fang, Science {\bf 329}, 61 (2010).
\bibitem{science340} 
C.-Z. Chang, {\it et. al.} Science {\bf 340}, 167 (2013).
\bibitem{nature438} 
H. Zhang, C.-X. Liu, X.-L. Qi, X. Dai, Z. Fang, and S.-C. Zhang, Nat. Phys. {\bf 5}, 438 (2009).
\bibitem{nature452} 
D. Hsieh, D. Qian, L. Wray, Y. Xia, Y. S. Hor, R. J. Cava, and M. Z. Hasan, Nature {\bf 452}, 970 (2008).
\bibitem{Hassan} 
M. Z. Hasan and C. L. Kane, Rev. Mod. Phys. {\bf 82}, 3045 (2010).
\bibitem{prl983D} 
L. Fu, C. L. Kane and E. J. Mele,  Phys. Rev. Lett. {\bf 98}, 106803 (2007).
\bibitem{prb76fu} 
L. Fu and C. L. Kane, Phys. Rev. B. {\bf 76}, 045302 (2007).
\bibitem{science325} 
Y. L. Chen, {\it et. al.} Science, {\bf 325}, 178 (2009).
\bibitem{nature584} 
Y. Zhang, {\it et. al.} Nat. Phys. {\bf 6}, 584 (2010).
\bibitem{prb81041307} 
C. X. Liu, H. J. Zhang, B. Yan, X. L. Qi, T. Frauenheim, X. Dai, Z. Fang, S. C. Zhang, Phys. Rev. B. {\bf 81}, 041307(R) (2010).
\bibitem{electrically} 
J. Wang, B. Lian and S.-C. Zhang, Phys. Rev. Lett. {\bf 115}, 036805 (2015).
\bibitem{prl113137201}
 X. Kou, {\it et. al.} Phys. Rev. Lett. {\bf 113}, 137201 (2014).
\bibitem{nature10731} 
J. G. Checkelsky, {\it et. al.} Nat. Phys. {\bf 10}, 731 (2014).
\bibitem{prb81H} 
H-Z. Lu, W-Y. Shan, W. Yao, Q. Niu, and S-Q. Shen,  Phys. Rev. B. {\bf 81}, 115407 (2010)
\bibitem{prl2013QT} 
H-Z. Lu, A. Zhao and S-Q. Shen, Phys. Rev. Lett. {\bf 111}, 146802 (2013).
\bibitem{wang}
J. Wang , B. Lian , H. Zhang  and S. C. Zhang , Phys. Rev. Lett. {\bf 111}, 086803 (2013) .
\bibitem{universality}
J. Wang, B. Lian and S-C. Zhang, Phys. Rev. B. {\bf 89}, 085106 (2014).
\bibitem{epl2014}
S-G. Cheng, Euro. Phys. J {\bf 105}, 57004 (2014).
\bibitem{Grauer}
S. Grauer, {\it et. al.} Phys. Rev. B. {\bf 92}, 201304 (2015).
\bibitem{Shiranzaee1}
M. Shiranzaei, F. Parhizgar, J. Fransson, H. Cheraghchi, Phys. Rev. B. {\bf 95}, 235429 (2017).
\bibitem{Shiranzaee2}
M. Shiranzaei, H. Cheraghchi, F. Parhizgar, Phys. Rev. B.  {\bf 96}, 024413 (2017).
\bibitem{transport_1}
L. He, {\it et. al.}, Nano Lett., {\bf 12} 1486 (2012).
\bibitem{transport_2}
T. Zhang, N. Levy, J. Ha, Y. Kuk and J. A. Stroscio, Phys. Rev. B., {\bf 87}, 115410 (2013).
\bibitem{transport_3}
E. I. Rogacheva, {\it et. al.}, Thin Solid Films, {\bf 594}, 109 (2015). 
\bibitem{transportTI_nanoscale}
H. Li, J. M. Shao, H. B. Zhang and G. W. Yang, Nanoscale, {\bf 6}, 3127 (2014).
\bibitem{rotational}
A.A. Farajian, K. Esfarjani  and Y. Kawazoe, Phys. Rev. Lett. {\bf 82}, 5084 (1999).
\bibitem{cheraghchi_evenZGNR}
H. Cheraghchi, H. Esmailzade, Nanotechnology, {\bf 21}, 205306  (2010).
\bibitem{cheraghchi_oddZGNR}
H. Cheraghchi, Phys. Scr. {\bf 84}, 015702 (2011).
\bibitem{parity}
A. Cresti, G. Grosso and G. P. Parravicini, Phys. Rev. B. {\bf 78}, 115433 ( 2008).
\bibitem{parity1}
A. Cresti, G. Grosso and G. P. Parravicini, Phys. Rev. B. {\bf 77}, 233402 (2008).
\bibitem{cheraghchi_nanojunction}
H. Cheraghchi, K. Esfarjani, Phys. Rev. B. {\bf 78}, 085123 (2008).
\bibitem{effective} 
W.-Y. Shan, H.-Z. Lu and S.-Q. Shen, New J. Phys {\bf 12}, 043048 (2010).
\bibitem{Magnetic}
V. Kul bachinskii, {\it et. al.} JETP Letters, {\bf 73}, 352 (2001).
\bibitem{PRL2013_zhang}
J. Wang, B. Lian, H. Zhang, Y. Xu, and S-C. Zhang, Phys. Rev. Lett. {\bf 111}, 136801 (2013) .
\bibitem{geometrical}
D. Sticlet, F. Piechon, J-N. Fuchs, P. Kalugin, and P. Simon, Phys. Rev. B. {\bf 85}, 165456  (2012).
\bibitem{chiral_TI}
P. Hosur, S. Ryu, and A. Vishwanath, Phys. Rev. B. {\bf 81}, 045120 (2010).
\bibitem{Phys_scri_2015}
J. Wang, B. Lian and S-C. Zhang, Phys. Scr. T164, 014003  (2015). 
\bibitem{graphene_dolfus}
V. Nam Do, V. Hung Nguyen, P. Dollfus, and A. Bournel, J. Appl. Phys. {\bf 104}, 063708 (2008).
\bibitem{buttiker}
Electronic Transport in Mesoscopic Systems, edited by S. Datta (Cambridge University Press, Cambridge, UK, 1995).
\bibitem{lopez}
M. P. L. Sancho, J. M. L. Sancho and J. Rubio, J. Phys. F. {\bf 15}, 851 (1985).
\bibitem{Esaki}
L. Esaki, Phys. Rev {\bf 109} , 603 (1958). 




\end{thebibliography}
\end{document}